\documentclass[aps,prl,reprint,superscriptaddress,longbibliography,nofootinbib,floatfix]{revtex4-2}

\usepackage{graphicx}
\usepackage{amsmath,amssymb,bm}
\usepackage{booktabs,array,multirow}
\usepackage{microtype}
\usepackage{placeins}
\usepackage{hyperref}
\hypersetup{colorlinks=true,linkcolor=blue,citecolor=blue,urlcolor=blue}
\graphicspath{{./}{../}}

\begin{document}

\title{Charge Symmetry Beyond Space-Group Equivalence}

\author{Qiu-Shi Huang}
\affiliation{Eastern Institute of Technology, Ningbo 315200, China}
\author{Xin-Gao Gong}
\affiliation{Key Laboratory for Computational Physical Sciences (MOE), State Key Laboratory of Surface Physics, Department of Physics, Fudan University, Shanghai 200433, China}
\author{Su-Huai Wei}
\email{suhuaiwei@eitech.edu.cn}
\affiliation{Eastern Institute of Technology, Ningbo 315200, China}

\date{August 15, 2026}

\begin{abstract}
Crystallographic space-group symmetry \(G_{\rm lat}\), determined by atomic species and their spatial arrangement, is one of the most important descriptors in solid-state physics, underlying the classification of electronic states, spectral degeneracies, order parameters, and phase transitions. Yet the symmetry \(G\) of a crystal also depends on the electronic coupling network between atomic sites, including electron hopping, Coulomb interactions, and orbital hybridization. This raises a fundamental question: must \(G\) reproduce every equivalence relation imposed by \(G_{\rm lat}\)? Equivalently, must symmetry-related atoms at the same Wyckoff position be electronically identical, while atoms at inequivalent Wyckoff positions are electronically distinct? We develop a systematic theory of interaction-controlled electronic equivalence, with site charge imbalance as an order parameter whose stability is governed by the competition between onsite charging cost and intersite Coulomb gain. Group-theoretical analysis identifies the site-exchange operations lost from or added to \(G_{\rm lat}\). Sites identified as equivalent by \(G_{\rm lat}\) can spontaneously develop charge imbalance, lowering the realized symmetry to \(G\subset G_{\rm lat}\). Conversely, sites identified as inequivalent by \(G_{\rm lat}\) can remain equivalent through a hidden low-energy gauge symmetry. Within the low-energy \((s,p_z)\) manifold, this realizes \(G\supset G_{\rm lat}\) and protects near-Fermi degeneracies that appear accidental in a \(G_{\rm lat}\)-based analysis. First-principles calculations verify both scenarios and establish pressure as a control parameter: it destabilizes the charge-equivalent state in Type I, whereas in Type II it destroys the hidden equivalence, splits the near-Fermi doublets, and can drive a metal-insulator transition.
\end{abstract}

\maketitle

Crystallographic space-group symmetry \(G_{\rm lat}\) is one of the organizing principles of solid-state physics. It is determined by atomic species and their spatial arrangement, defines site equivalence through Wyckoff positions, and underlies the classification of electronic bands, symmetry-enforced degeneracies, topological indicators, phase transitions, and order parameters \cite{ref34,ref35,ref36,ref37,ref38,ref11,bradlyn2016fermions,bradlyn2017tqc}. Atoms on the same Wyckoff position are related by operations of \(G_{\rm lat}\), whereas atoms on inequivalent Wyckoff positions are not. This leads to a natural expectation. The former should be electronically identical and the latter electronically distinct.

The symmetry \(G\) realized by a crystal, however, also depends on the electronic interaction network between its atomic sites, including electron hopping, Coulomb interactions, and orbital hybridization. In the usual situation, \(G\) reproduces the equivalence relations imposed by \(G_{\rm lat}\). For a nondegenerate ground state, observables such as the charge density normally inherit the crystal symmetry, consistent with Neumann's principle \cite{ref39}. Two departures are fundamental. A degenerate manifold can support spontaneous symmetry breaking, so that the selected state retains only a subgroup of the symmetry of the electronic problem \cite{ref31,ref32}. Conversely, projection onto a restricted low-energy manifold can reveal a hidden site-exchange operation absent from the atomic structure \cite{zhang2023knonsymmorphic}. The site-equivalence relations realized by the electronic problem therefore need not coincide with those imposed by the atomic lattice.

We show that these departures occur in two opposite directions. An interaction-driven instability can remove a site-exchange operation contained in \(G_{\rm lat}\), producing \(G\subset G_{\rm lat}\). Conversely, a gauge-conjugated site exchange can add an equivalence operation absent from the atomic space group, producing \(G\supset G_{\rm lat}\) within the relevant low-energy manifold. The first reconstruction splits one crystallographic orbit into two electronic sublattices. The second joins two crystallographic orbits into one low-energy equivalence class. Together, they show that the equivalence structure realized by a crystal is determined jointly by atomic geometry and the electronic interaction network.

We develop a systematic theory of interaction-controlled electronic equivalence. Site charge imbalance serves as an order parameter whose stability is governed by the competition between onsite charging cost and intersite Coulomb gain, while group-theoretical analysis identifies the site-exchange operations lost from or added to \(G_{\rm lat}\). First-principles calculations across elemental crystals verify both symmetry reconstructions. Pressure provides a common control parameter. It destabilizes the charge-equivalent state in Type I, whereas in Type II it destroys a pre-existing hidden equivalence, splits the near-Fermi doublets, and can drive a metal--insulator transition.

Recognizing that \(G\) need not coincide with \(G_{\rm lat}\) changes how the symmetry and electronic structure of crystals should be calculated, classified, and understood. Enforcing \(G_{\rm lat}\) on the electronic degrees of freedom in a first-principles calculation can exclude a lower-symmetry electronic ground state even when the atomic framework remains unchanged. Conversely, classifying bands by \(G_{\rm lat}\) alone can misidentify degeneracies protected by a hidden low-energy symmetry as accidental \cite{ref33}. Hidden equivalence can also allow crystallographically distinct structures to share nearly identical low-energy spectra, providing a microscopic route to structural competition and frustration \cite{ref10}. This perspective offers an organizing principle for delayed metallization and persistent insulating behavior in hydrogen, transparent or insulating phases in compressed alkali metals, and the widespread appearance of complex low-symmetry structures in nominally simple elements \cite{ref2,ref3,ref4,ref5,ref6,ref8,ref9,ref13,ref21,ref22,ref23,ref24,ref25,ref29,ref40,ref41,ref42,ref43,ref44,ref45}. Pressure provides a particularly effective means of revealing these effects by continuously tuning intersite interactions without changing chemical composition. Electronic equivalence should therefore be regarded as an interaction-dependent structure of a crystal rather than as a fixed consequence of its atomic space group.

To make the distinction between space-group-imposed site equivalence and electronic equivalence explicit, we formulate a minimal Landau theory for the loss of site-exchange equivalence through charge transfer. The theory isolates the general energetic mechanism by which a charge-equivalent state develops a finite site imbalance. Its coefficients can be tuned by any parameter that changes the balance of electronic interactions. Pressure is used in the examples below because it continuously changes intersite distances without changing chemical composition.

Consider two sites \(A\) and \(B\), and define the charge-transfer order parameter \(\delta\) through
\begin{equation}
Q_A=Q_0+\delta e,\qquad Q_B=Q_0-\delta e,
\label{eq:charge-order}
\end{equation}
Overall charge neutrality is preserved, and the charge-equivalent state corresponds to \(\delta=0\). Whenever an explicit or hidden site-exchange symmetry relates \(A\) and \(B\), it sends \(\delta\) to \(-\delta\). Invariance under this exchange requires \(F(\delta)=F(-\delta)\). The charge imbalance is therefore odd under site exchange, while the free energy contains only even powers of \(\delta\).

The central point is that electronic equivalence is an energetic condition rather than a purely crystallographic label. Near the charge-equivalent state, the lowest-order bounded free energy is
\begin{equation}
F_{\rm cell}(\delta,P)=A_2(P)\delta^2+A_4(P)\delta^4,
\qquad A_4(P)>0,
\label{eq:landau}
\end{equation}
with
\begin{equation}
A_2(P)=A_{\rm intra}(P)-A_{\rm inter}(P).
\label{eq:a2}
\end{equation}
Here \(A_{\rm intra}\) is the onsite charging cost required to create a local charge deviation, while \(A_{\rm inter}>0\) is the intersite energy gained by arranging opposite charge deviations on neighboring sites. The charge-equivalent state is stable for \(A_2(P)>0\). Its quadratic stiffness vanishes at \(A_2(P_c)=0\). For \(A_2(P)<0\), the charge-equivalent state becomes unstable and a pair of symmetry-related charge-transfer states emerges,
\begin{equation}
\delta=\pm\sqrt{\frac{-A_2(P)}{2A_4(P)}}.
\label{eq:delta-minimum}
\end{equation}

\begin{figure}[tb]
\centering
\includegraphics[width=\linewidth]{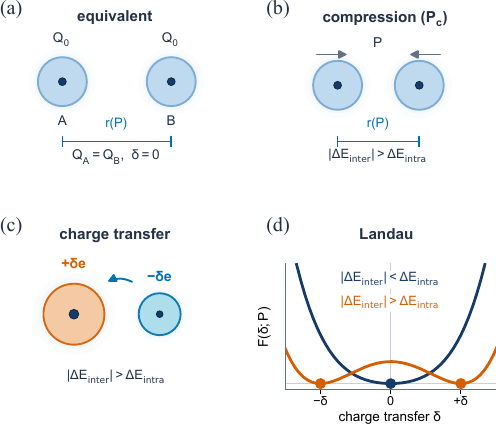}
\caption{Interaction-driven charge-transfer instability. (a) Charge-equivalent state. (b,c) Reducing the separation \(r\) enhances the intersite Coulomb gain until \(|\Delta E_{\rm inter}|>\Delta E_{\rm intra}\). (d) The \(\delta=0\) minimum then bifurcates into two symmetry-related charge-transfer states.}
\label{fig:model}
\end{figure}

This instability is illustrated by the free-energy evolution in Fig.~\ref{fig:model}.

The Landau condition can be converted into a transparent geometric criterion. Treating the outer valence region on each site as a charge cloud of radius \(R\) gives \(A_{\rm intra}\sim ke^2/R\). For an alternating charge pattern, the Madelung-enhanced intersite gain scales as \(A_{\rm inter}\sim ke^2M(P)/r(P)\), where \(r(P)\) is the relevant intersite separation and \(M(P)\) is the Madelung constant of the specified charge pattern. Comparing the two energy scales gives the central result
\begin{equation}
\boxed{\frac{M(P)}{r(P)}>\frac{1}{R}
\quad\Longleftrightarrow\quad
r(P)<M(P)R.}
\label{eq:criterion}
\end{equation}
This relation is the central working criterion of the materials analysis. It reduces the onset of charge transfer to a transparent structural condition. A small intersite separation strengthens the intersite Coulomb gain. A large Madelung constant amplifies the collective lattice contribution. A large charging radius reduces the onsite charging cost. Corrections beyond the monopole approximation renormalize the coefficients but preserve the \(\delta\leftrightarrow-\delta\) symmetry and the form of the instability. The electrostatic derivation is given in End Matter Appendix A. The numerical implementation is described in the Supplemental Material \cite{supplemental}.

The same Landau instability acts on two different sources of charge equivalence. In Type I, charge equivalence is imposed by a site-exchange operation of \(G_{\rm lat}\). A finite \(\delta\) spontaneously breaks this lattice-imposed equivalence and produces \(G\subset G_{\rm lat}\). In Type II, charge equivalence is protected by an additional hidden symmetry of a restricted low-energy manifold. A finite \(\delta\) breaks this hidden equivalence and removes the symmetry extension \(G\supset G_{\rm lat}\).

\paragraph{Type I: \(G\subset G_{\rm lat}\).}
Type I realizes \(G\subset G_{\rm lat}\) through spontaneous electronic symmetry breaking. Consider a crystal with a single space-group orbit. Crystallography assigns all sites to one equivalence class, apparently leaving no distinction that could support site-selective charge transfer. Nevertheless, the electronic state can break a site-exchange operation that remains present in the atomic lattice. Once the charge-transfer channel becomes energetically favorable, the system selects one of a pair of symmetry-related charge configurations while the atomic framework remains unchanged.

The group reduction is explicit in BCC. Its atomic space group is \(G_{\rm lat}=Im\bar3m\), and the body-centering translation
\begin{equation}
t_c=\left\{E\,\middle|\,\tfrac12,\tfrac12,\tfrac12\right\}
\in G_{\rm lat}
\label{eq:tc}
\end{equation}
exchanges the two sublattices \(A\) and \(B\). For the charge-transfer order parameter \(\Delta Q=Q_A-Q_B\), this operation gives
\begin{equation}
t_c:\Delta Q\longmapsto-\Delta Q.
\label{eq:tc-order}
\end{equation}
The charge-equivalent state \(\Delta Q=0\) preserves \(t_c\). When the quadratic Landau coefficient changes sign, the two states \(\Delta Q=\pm\Delta Q_0\) become degenerate. Either selected state loses \(t_c\) and realizes
\begin{equation}
G=Pm\bar3m\subset G_{\rm lat}=Im\bar3m,
\label{eq:typeIgroup}
\end{equation}
while the atomic BCC framework remains fixed. The two charge-transfer domains are exchanged by \(t_c\), and the crystallographic group decomposes as
\begin{equation}
G_{\rm lat}=G\cup t_cG.
\label{eq:typeIcoset}
\end{equation}
The atomic framework therefore retains \(G_{\rm lat}\), while either selected electronic state retains only \(G\). A single crystallographic orbit is spontaneously split into two electronic sublattices without moving the nuclei.

To realize this single-orbit instability in real materials, we seek a lattice geometry that maximizes the electrostatic tendency toward charge transfer. The BCC lattice is ideal in this respect. It is one of the simplest and highest-symmetry metallic structures, with all atoms forming a single space-group orbit in \(Im\bar3m\). When BCC is viewed as two interpenetrating sublattices, its alternating charge-transfer pattern has the same electrostatic lattice sum as the CsCl structure. This gives the comparatively large Madelung constant \(M_{\rm BCC}=1.762675\) \cite{ref16,ref26,ref27}. According to Eq.~\eqref{eq:criterion}, this large Madelung constant makes BCC particularly favorable for a charge-transfer instability \cite{ref1}.

The alkali metals provide a systematic chemical series for testing this prediction. Their diffuse single-valence-\(s\) states give large charging radii, while the variation of atomic size across Li, Na, K, Rb, and Cs provides a controlled test of the predicted scaling with \(R\) \cite{ref14,ref15}. Using empirical covalent radii as estimates of \(R\), the geometric criterion predicts \(r_c=M_{\rm BCC}R\). A larger \(M_{\rm BCC}\) or \(R\) moves the instability to a larger interatomic distance and therefore reduces the compression required to reach it. We compare different elements using the reduced separation
\begin{equation}
f\equiv\frac{r}{M_{\rm BCC}R}.
\label{eq:reduced-separation}
\end{equation}
The geometric instability is expected near \(f=1\).

For each alkali element, we performed all-electron first-principles calculations using WIEN2k \cite{ref17,ref18} and compared two electronic states at the same fixed BCC atomic geometry. The symmetry-constrained state preserves the full \(Im\bar3m\) equivalence. The symmetry-free two-sublattice state allows the electronic degrees of freedom to relax without imposing this equivalence. The atomic BCC framework is identical in the two calculations. Folded-equivalent physical \(k\)-point meshes ensure identical Brillouin-zone sampling between the one-site and two-site descriptions. The convergence, charge-partition, and common onset protocols are provided in the Supplemental Material \cite{supplemental}.

Figure~\ref{fig:bcc} summarizes the results across the alkali-metal series. At large \(f\), the symmetry-free calculation returns the charge-equivalent state, and both \(E_{\rm sym}-E_{\rm free}\) and \(|\Delta Q|\) remain essentially zero. As \(f\) approaches and falls below unity, a positive energy gain and a finite charge imbalance develop together for Li, Na, K, Rb, and Cs. The calculated onset distances agree with the geometric prediction \(r_c=M_{\rm BCC}R\) to within approximately \(6\%\), as summarized in Table S1 \cite{supplemental}. This agreement establishes the dimensionless ratio \(f\) as a transferable organizing scale for Type-I electronic symmetry breaking across chemically distinct elemental crystals.

\begin{figure}[tb]
\centering
\includegraphics[width=\linewidth]{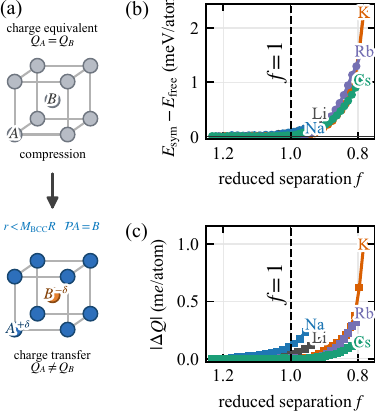}
\caption{Type-I symmetry breaking in BCC alkali metals. (a) The charge-transfer state loses \(t_c\) while the atomic BCC framework is fixed. (b) Energy gain of the symmetry-free branch and (c) site-projected charge imbalance versus \(f=r/(M_{\rm BCC}R)\). The dashed line marks \(f=1\).}
\label{fig:bcc}
\end{figure}

These calculations directly verify the group reduction in Eq.~\eqref{eq:typeIgroup}. The body-centering translation remains an operation of the fixed atomic framework, but it is absent from either selected charge-transfer state. This electronic and atomic symmetry mismatch also creates a natural coupling to the lattice. Once the electronic state has lost an operation retained by the atomic framework, a lattice distortion with the same symmetry reduction becomes compatible with the electronic order and can lower the coupled electron--lattice energy. The Type-I instability therefore provides a microscopic route by which simple high-symmetry phases can become susceptible to the complex and low-symmetry structures widely observed in compressed elemental solids \cite{ref2,ref3,ref4,ref5,ref7,ref29,ref30}.

\begin{figure}[!tb]
\centering
\includegraphics[width=\linewidth]{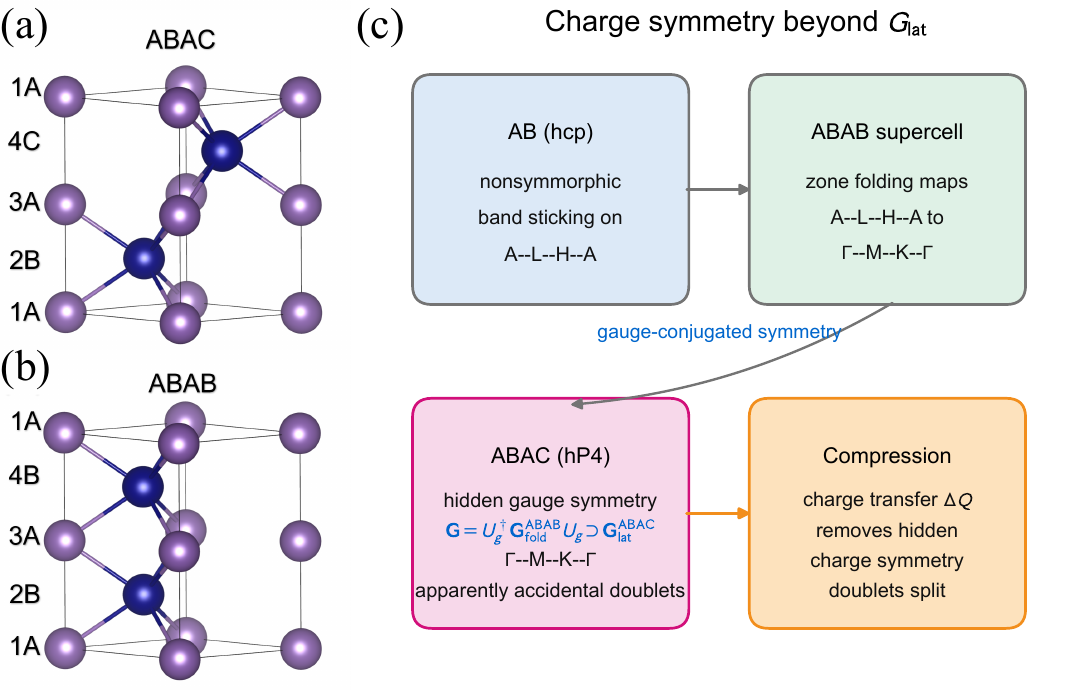}
\caption{Type-II charge symmetry beyond \(G_{\rm lat}\). (a,b) ABAC and ABAB stackings. (c) Nonsymmorphic band sticking in the AB parent folds to the ABAB basal path. Gauge conjugation transports this symmetry to the low-energy ABAC manifold, where \(\widetilde g\) joins the inequivalent sites. Compression-induced charge transfer removes this hidden equivalence.}
\label{fig:typeII-mechanism}
\end{figure}

\paragraph{Type II: \(G\supset G_{\rm lat}\).}
Type II occurs when a restricted low-energy manifold contains an additional site-exchange symmetry absent from the atomic space group. This hidden operation connects the projected electronic sectors associated with crystallographically inequivalent sites and joins them into one electronic equivalence class. The symmetry realized by the electronic problem can therefore exceed that defined by the atomic positions.

Close-packed alkali structures provide a clear realization of this symmetry extension through three linked anomalies \cite{ref28}. In ABAC (hP4), \(G_{\rm lat}^{\rm ABAC}=P6_3/mmc\) has the inequivalent representatives \(X_1=(0,0,0)\) and \(X_2=(1/3,2/3,1/4)\). Nevertheless, \textit{(i)} their calculated charges remain equal at low pressure, \textit{(ii)} the near-Fermi bands form extended doublets along K--\(\Gamma\)--M even though the paired branches belong to different one-dimensional irreducible representations of \(G_{\rm lat}^{\rm ABAC}\), and \textit{(iii)} crystallographically distinct ABAC and ABAB stackings exhibit nearly identical low-energy dispersions. These observations are three manifestations of one hidden electronic equivalence. The site-resolved analysis and direct AB/ABAB/ABAC comparison are provided in the Supplemental Material and Fig. S1 \cite{supplemental}.

Their common origin is summarized in Fig.~\ref{fig:typeII-mechanism}. In the folded four-layer ABAB reference structure, the operation
\begin{equation}
g=\left\{-I\,\middle|\,\tfrac13,\tfrac23,\tfrac14\right\}
\in G_{\rm lat}^{\rm ABAB}
\label{eq:g}
\end{equation}
is inversion followed by a fractional translation. It exchanges \(X_1\) and \(X_2\), modulo a lattice vector, so their charge equivalence is explicitly enforced by the ABAB space group. In ABAC, the fourth layer occupies the \(C\), rather than \(B\), position. The same real-space operation is therefore absent from \(G_{\rm lat}^{\rm ABAC}\). Consequently, the observed equivalence between \(X_1\) and \(X_2\) cannot be explained by the ABAC space group alone and lies beyond the conventional expectation that inequivalent Wyckoff sites must be electronically distinct. It signals an additional symmetry of the projected low-energy problem.

The hidden gauge equivalence of close-packed structures was previously established in an \(s\)-orbital model \cite{ref10}. We generalize it to the coupled \((s,p_z)\) manifold and identify this gauge symmetry as the origin of the near-Fermi doublets that appear accidental in an ABAC space-group analysis. In the short-range four-layer model, reversing a stacking chirality changes the \((s,p_z)\) interlayer block only by a Bloch phase. Because ABAB and ABAC have the same net stacking chirality, these phases are removed by the periodic layer gauge
\begin{equation}
\begin{aligned}
U_g(k)&=\operatorname{diag}\left(1,1,1,e^{-ik\cdot s}\right)\otimes I_2,\\
U_g^\dagger(k)H_{\rm ABAB}(k)U_g(k)&=H_{\rm ABAC}(k).
\end{aligned}
\label{eq:typeIIUg}
\end{equation}
Here \(s\) is the in-plane shift between opposite stacking steps and \(I_2\) acts on the \((s,p_z)\) orbital subspace. Equation~\eqref{eq:typeIIUg} establishes the unitary equivalence of ABAB and ABAC within this low-energy manifold and transports the explicit ABAB site exchange into the ABAC electronic problem.

Let \(D_g(k)\) denote the Bloch representation of the ABAB operation \(g\). Its gauge-conjugated image in ABAC is
\begin{equation}
\begin{aligned}
\widetilde g(k)&=U_g^\dagger(gk)D_g(k)U_g(k),\\
\widetilde g(k)H_{\rm ABAC}(k)\widetilde g^\dagger(k)&=H_{\rm ABAC}(gk).
\end{aligned}
\label{eq:typeIIhiddenop}
\end{equation}
Thus \(\widetilde g\) exchanges the projected \(X_1\) and \(X_2\) sectors and enforces their electronic equivalence, although no corresponding real-space operation belongs to \(G_{\rm lat}^{\rm ABAC}\). Gauge conjugation of the folded ABAB symmetry representation gives
\begin{equation}
G=U_g^\dagger G_{\rm fold}^{\rm ABAB}U_g
\supset G_{\rm lat}^{\rm ABAC},
\qquad
\widetilde g\in G\setminus G_{\rm lat}^{\rm ABAC}.
\label{eq:typeIIgroup}
\end{equation}
Here \(G_{\rm fold}^{\rm ABAB}\) is projected onto the same low-energy manifold. The complete layer-by-layer derivation of Eqs.~\eqref{eq:typeIIUg}--\eqref{eq:typeIIgroup} is given in End Matter Appendix B and the Supplemental Material \cite{supplemental}.

The apparently accidental ABAC doublets follow from three linked steps. \textit{(i)} In the AB parent, the antiunitary operation \(\Theta=\{C_{2z}|c/2\}\mathcal T\) obeys \(\Theta^2=e^{-ik_zc}=-1\) on \(k_z=\pi/c\), enforcing twofold band sticking along A--L--H--A \cite{young2015dirac,parameswaran2013nonsymmorphic,wieder2018wallpaper}. \textit{(ii)} Doubling the cell to ABAB maps \(A\to\Gamma\), \(L\to M\), and \(H\to K\), so this boundary degeneracy folds onto \(\Gamma\)--M--K--\(\Gamma\). \textit{(iii)} Equation~\eqref{eq:typeIIUg} transports the folded spectrum and its symmetry generators into the ABAC low-energy manifold, where the doublets are inherited together with \(\widetilde g\).

Thus the ABAC doublets are not accidental in the low-energy electronic problem. They appear accidental only when classified by \(G_{\rm lat}^{\rm ABAC}\) alone. Figure S1 \cite{supplemental} displays the full AB--ABAB--ABAC sequence. This construction connects close-packing gauge equivalence to latent-symmetry-protected degeneracies \cite{rontgen2021,ref47}.

\begin{figure}[!tb]
\centering
\includegraphics[width=\linewidth]{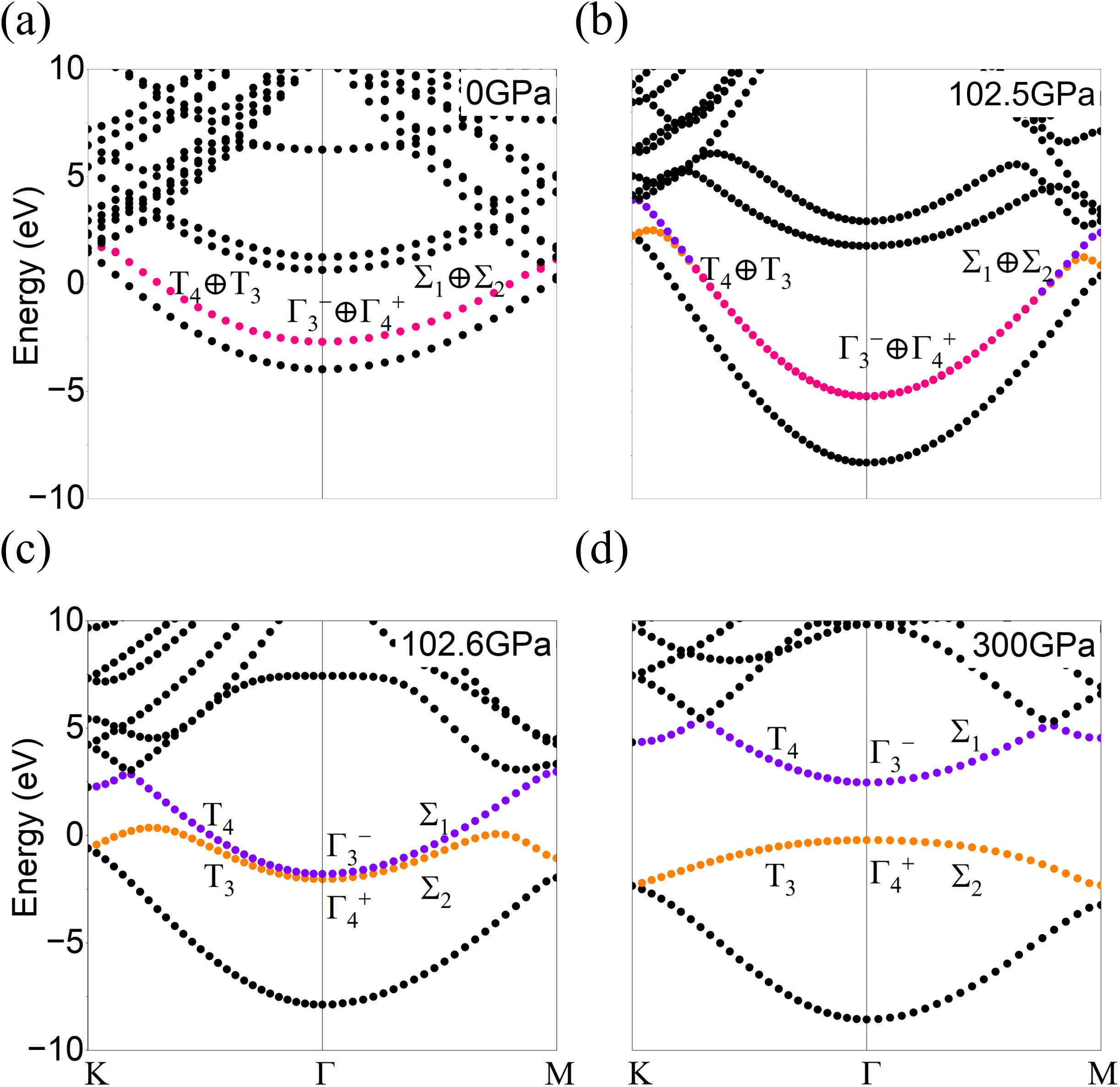}
\caption{Pressure evolution of the near-Fermi ABAC doublets. (a) At 0 GPa, the hidden-symmetry-related pairs remain nearly degenerate. (b) The same doublets persist immediately below the transition at 102.5 GPa. (c) At 102.6 GPa, charge transfer splits the partners. (d) At 300 GPa, the splitting is fully developed. Magenta denotes the hidden-symmetry-related pairs before charge transfer, whereas purple and orange track the split partners.}
\label{fig:hP4-bands}
\end{figure}

Compression removes this hidden extension in two connected ways. Higher \(p_{x,y}\) and \(d\) admixture and longer-range hopping violate the short-range \((s,p_z)\) gauge relation, while charge transfer produces a finite \(X_1/X_2\) onsite imbalance and removes \(\widetilde g\). Figure~\ref{fig:hP4-bands} shows the resulting splitting of the \(T_4\oplus T_3\), \(\Gamma_3^-\oplus\Gamma_4^+\), and \(\Sigma_1\oplus\Sigma_2\) partners across K--\(\Gamma\)--M. Since each pair belongs to different one-dimensional irreducible representations, the splitting is a direct energetic separation. Near \(\Gamma\), it separates the valence and conduction edges and opens an insulating gap.

The group construction explains why the low-pressure equivalence exists, while the charge-transfer criterion determines when it is lost. In ABAC, the onset remains governed by \(r_c=M_{\rm hP4}R\), with the geometry-dependent Madelung constant evaluated at the corresponding \(c/a\). Across hP4 Li, Na, and K, the calculated transition distances follow this prediction to within about \(7\%\) (Table S2 \cite{supplemental}). The same energetic competition therefore controls both the loss of lattice-imposed equivalence in Type I and the destruction of hidden equivalence in Type II.

\FloatBarrier

The two types reveal a single principle: the electronic interaction network can reconstruct the equivalence relations implied by atomic geometry. In Type I, an interaction-driven instability removes a lattice-imposed operation and splits one crystallographic orbit into two electronic sublattices. In Type II, a gauge-conjugated operation absent from the atomic space group joins two crystallographic orbits within a restricted low-energy manifold. The alkali-metal series demonstrates that neither reconstruction is tied to a single element, while the common radius criterion shows that both are governed by a transferable interaction scale.

This distinction changes how both calculations and spectra should be interpreted. Enforcing the full atomic space group can prevent a first-principles calculation from reaching a lower-symmetry electronic state. Conversely, classifying bands only by the atomic space group can obscure a hidden symmetry and misidentify protected degeneracies as accidental. The resulting electronic--lattice mismatch also supplies a microscopic route through which electronic order can promote structural reconstruction. Atomic positions establish the crystallographic baseline, but interactions determine whether the symmetry realized by the electronic problem coincides with, falls below, or rises above it. Electronic interactions can therefore reconstruct the equivalence relations on which symmetry classification itself is built.

\begin{acknowledgments}
This work was supported by the National Key Research and Development Program of China (Grant No. 2024YFA1409800) and the National Natural Science Foundation of China (Grant No. 12688201).
\end{acknowledgments}

\section*{Data Availability}
The processed data supporting the figures and tables are contained in the manuscript and Supplemental Material. The underlying first-principles input and output files are available from the corresponding author upon reasonable request.

\bibliography{references}

\section*{End Matter Appendix A: Charge-transfer criterion}

Let the two site charges be \(Q_A=Q_0+\delta e\) and \(Q_B=Q_0-\delta e\). A site-exchange operation sends \(\delta\) to \(-\delta\), so the energy measured from the charge-equivalent state is even in \(\delta\). To quartic order,
\begin{equation}
\Delta E(\delta)=A_2\delta^2+A_4\delta^4,
\qquad A_4>0.
\label{eq:em-landau}
\end{equation}
The charge-equivalent state is stable for \(A_2>0\). For \(A_2<0\), the two minima are
\begin{equation}
\delta_0=\pm\sqrt{-\frac{A_2}{2A_4}},
\qquad
\Delta E(\delta_0)=-\frac{A_2^2}{4A_4}.
\label{eq:em-minima}
\end{equation}

The quadratic coefficient contains the local charging cost and the electrostatic energy gained by arranging opposite charge deviations on neighboring sublattices. Approximating each outer valence cloud by a spherical shell of radius \(R\), the self-energy of charge \(Q\) is \(kQ^2/(2R)\), where \(k=(4\pi\varepsilon_0)^{-1}\). The two-site contribution is
\begin{equation}
\Delta E_{\rm intra}
=2\frac{k(\delta e)^2}{2R}
=\frac{ke^2}{R}\delta^2.
\label{eq:em-self}
\end{equation}
For an alternating charge pattern, define the Madelung constant \(M\) by the lattice energy per two-site cell,
\begin{equation}
\Delta E_{\rm inter}
=-\frac{ke^2M}{r}\delta^2.
\label{eq:em-inter}
\end{equation}
Here \(r\) is the nearest separation between opposite charge deviations. Combining Eqs.~\eqref{eq:em-self} and \eqref{eq:em-inter} gives
\begin{equation}
A_2\simeq ke^2\left(\frac1R-\frac Mr\right),
\label{eq:em-a2}
\end{equation}
and hence
\begin{equation}
A_2<0
\quad\Longleftrightarrow\quad
r<MR.
\label{eq:em-criterion}
\end{equation}
Equation~\eqref{eq:em-criterion} is Eq.~\eqref{eq:criterion} of the main text. Multipole terms, screening, and the pressure dependence of the valence cloud renormalize \(A_2\), \(A_4\), \(M\), and \(R\). They preserve the transformation \(\delta\mapsto-\delta\) and therefore do not change the form of the instability.

For BCC, \(r=\sqrt3a/2\). The body-centering translation
\begin{equation}
t_c=\left\{E\,\middle|\,\tfrac12,\tfrac12,\tfrac12\right\}
\label{eq:em-tc}
\end{equation}
exchanges the two charge sublattices and sends \(\delta\) to \(-\delta\). A state with \(\delta\ne0\) does not retain \(t_c\). The operations that preserve either selected charge sublattice form \(G=Pm\bar3m\). Since \(t_c^2\in G\), the reduction has index two,
\begin{equation}
G_{\rm lat}=G\cup t_cG,
\qquad
Pm\bar3m\subset Im\bar3m.
\label{eq:em-coset}
\end{equation}
The two domains \(\delta=\pm\delta_0\) are degenerate and are exchanged by \(t_c\). The atomic BCC lattice retains \(G_{\rm lat}=Im\bar3m\), while either selected electronic state realizes \(G=Pm\bar3m\).

\section*{End Matter Appendix B: Low-energy hidden symmetry}

In the four-layer ABAB reference cell, define
\begin{equation}
\begin{aligned}
X_1&=(0,0,0), & X_2&=\left(\tfrac13,\tfrac23,\tfrac14\right),\\
X_1'&=\left(0,0,\tfrac12\right), &
X_2'&=\left(\tfrac13,\tfrac23,\tfrac34\right).
\end{aligned}
\label{eq:em-sites}
\end{equation}
The operation \(g\) in Eq.~\eqref{eq:g} acts as
\begin{equation}
\begin{aligned}
gX_1&=X_2, &\qquad gX_2&=X_1,\\
gX_1'&=X_2', & gX_2'&=X_1'
\end{aligned}
\qquad (\mathrm{mod}\ \mathbf R).
\label{eq:em-site-map}
\end{equation}
It places the two ABAB sublattices in one crystallographic orbit. In ABAC, the fourth-layer site is \((2/3,1/3,3/4)\), whereas \(gX_1'=(1/3,2/3,3/4)\). Thus \(g\notin G_{\rm lat}^{\rm ABAC}\).

Let a close-packed stacking be encoded by chiralities \(\sigma_j=\pm1\), with \(\sigma_j\) specifying the stacking step from layer \(j\) to layer \(j+1\). In the layer basis \(\Psi_k=(\psi_{1k},\ldots,\psi_{Mk})^T\), where \(\psi_{jk}=(c_{j,s,k},c_{j,p_z,k})^T\), the short-range Hamiltonian contains intralayer blocks \(D(k)\) and interlayer blocks \(T_{\sigma_j}(k)\). For a bond with out-of-plane direction cosine \(n\), the Slater--Koster block in the \((s,p_z)\) subspace is
\begin{equation}
M(n)=
\begin{pmatrix}
V_{ss\sigma} & nV_{sp\sigma}\\
-nV_{sp\sigma} &
n^2V_{pp\sigma}+(1-n^2)V_{pp\pi}
\end{pmatrix}.
\label{eq:em-sk}
\end{equation}
This block depends on \(n\) but not on the in-plane bond orientation. The two chiral interlayer-neighbor sets retained in the short-range model can therefore be paired as \(\eta_{+,\alpha}=\eta_{-,\alpha}+s\), where \(s_z=0\). Their Bloch sums obey
\begin{equation}
T_+(k)=\sum_\alpha e^{ik\cdot\eta_{+,\alpha}}M(n_\alpha)
=e^{ik\cdot s}T_-(k).
\label{eq:em-phase}
\end{equation}
Writing \(2\phi(k)=k\cdot s\), Eq.~\eqref{eq:em-phase} is equivalent to \(T_{\sigma_j}=e^{i\sigma_j\phi}\overline T\). For two sequences \(\sigma\) and \(\chi\), a diagonal layer gauge with phases \(\theta_j\) maps every interlayer block when
\begin{equation}
\theta_{j+1}-\theta_j
=\frac{\chi_j-\sigma_j}{2}\,k\cdot s.
\label{eq:em-layer-phase}
\end{equation}
Periodic boundary conditions require the sum of the phase increments to vanish modulo \(2\pi\). The condition is satisfied when \(\sum_j\sigma_j=\sum_j\chi_j\). The gauge
\begin{equation}
W_{\sigma,\chi}(k)=
\operatorname{diag}\left(e^{i\theta_1},\ldots,e^{i\theta_M}\right)
\otimes I_2
\label{eq:em-general-gauge}
\end{equation}
then obeys
\begin{equation}
W_{\sigma,\chi}^\dagger(k)H_\sigma(k)W_{\sigma,\chi}(k)=H_\chi(k).
\label{eq:em-isospectral}
\end{equation}
ABAB and ABAC have chirality sequences \((1,-1,1,-1)\) and \((1,-1,-1,1)\), both with zero net chirality. The choice \(\theta_1=\theta_2=\theta_3=0\) and \(\theta_4=-k\cdot s\) gives the explicit transformation \(U_g(k)\) in Eq.~\eqref{eq:typeIIUg}. Equation~\eqref{eq:em-isospectral} establishes their low-energy isospectrality.

Let \(D_g(k)\) be the ABAB Bloch representation of the site exchange in Eq.~\eqref{eq:g}. The ABAB symmetry relation and Eq.~\eqref{eq:em-isospectral} give
\begin{align}
\widetilde g(k)&=U_g^\dagger(gk)D_g(k)U_g(k),\nonumber\\
\widetilde g(k)H_{\rm ABAC}(k)\widetilde g^\dagger(k)
&=H_{\rm ABAC}(gk).
\label{eq:em-hidden-operation}
\end{align}
The operation \(\widetilde g\) exchanges the projected \(X_1\) and \(X_2\) sectors. It belongs to the symmetry of the projected ABAC Hamiltonian even though no corresponding real-space operation belongs to \(G_{\rm lat}^{\rm ABAC}\). Gauge conjugation of the folded ABAB representation therefore produces the extension in Eq.~\eqref{eq:typeIIgroup}.

The spectral consequence follows from the AB parent. The screw \(S_{2z}=\{C_{2z}|c/2\}\) and time reversal define \(\Theta=S_{2z}\mathcal T\). On \(k_z=\pi/c\),
\begin{equation}
\Theta^2=e^{-ik_zc}=-1,
\label{eq:em-theta}
\end{equation}
so \(|u_k\rangle\) and \(\Theta|u_k\rangle\) are orthogonal degenerate partners. This enforces the AB boundary doublets. Doubling the cell sends
\begin{equation}
A\mapsto\Gamma,\qquad L\mapsto M,\qquad H\mapsto K,
\label{eq:em-folding}
\end{equation}
so A--L--H--A folds onto \(\Gamma\)--M--K--\(\Gamma\). Equation~\eqref{eq:em-hidden-operation} transports the folded ABAB symmetry and its doublets to ABAC. The full layer Hamiltonian, the charge constraint generated by \(\widetilde g\), the Ewald definition of \(M\), the first-principles protocols, and the irrep-resolved alkali-metal results are provided in the Supplemental Material \cite{supplemental}.

\clearpage
\onecolumngrid
\setcounter{section}{0}
\setcounter{equation}{0}
\setcounter{figure}{0}
\setcounter{table}{0}
\renewcommand{\theequation}{S\arabic{equation}}
\renewcommand{\thefigure}{S\arabic{figure}}
\renewcommand{\thetable}{S\arabic{table}}
\renewcommand{\theHequation}{S.\arabic{equation}}
\renewcommand{\theHfigure}{S.\arabic{figure}}
\renewcommand{\theHtable}{S.\arabic{table}}
\begin{center}
{\large\bfseries Supplemental Material for\\[2pt]
Charge Symmetry Beyond Space-Group Equivalence}\par
\vspace{6pt}
Qiu-Shi Huang, Xin-Gao Gong, and Su-Huai Wei
\end{center}
\vspace{8pt}

\section{Scope of the Supplemental Material}

This Supplemental Material gives the derivations and numerical protocols used in the main text. The Type-I analysis identifies the crystallographic site exchange removed by a selected charge-transfer state. The Type-II analysis constructs the gauge-conjugated site exchange of the projected \((s,p_z)\) manifold. The first-principles data establish the corresponding trends across BCC and close-packed alkali-metal structures.

We use \(G_{\rm lat}\) for the crystallographic space group determined by atomic species and positions, and \(G\) for the symmetry realized by the interacting crystal at the scale under discussion. In Type II, \(G\) is understood within the projected low-energy \((s,p_z)\) manifold. The two types are
\begin{equation}
G\subset G_{\rm lat}\qquad\text{(Type I)}
\label{eq:si-two-classes-I}
\end{equation}
and
\begin{equation}
G\supset G_{\rm lat}\qquad\text{(Type II, within the low-energy manifold)}.
\label{eq:si-two-classes-II}
\end{equation}
The derivations below identify the operation removed or added in each case and connect it to the calculated charge and band observables.

\section{Computational and Numerical Protocols}

The primary all-electron calculations were performed with the full-potential (L)APW+lo implementation in WIEN2k \cite{ref17,ref18}, using the Perdew--Burke--Ernzerhof generalized-gradient approximation \cite{ref46}. Scalar-relativistic effects were included, while spin--orbit coupling was omitted so that the calculations and the spinless group analysis use the same symmetry setting. For the BCC calculations, hydrostatic compression was modeled by varying the cubic lattice constant while keeping the ions on the ideal BCC positions. The symmetry-constrained calculation enforces the full BCC equivalence, whereas the symmetry-unconstrained calculation uses the two-site cell and allows the electronic state to relax from an initially A/B-inequivalent density.

For the cross-element BCC results shown in Fig. 2 of the main text and Table~\ref{tab:si-bcc-onset}, the one-atom BCC cell and the two-atom symmetry-free cell were compared using exactly folded physical k-point meshes generated from an \(N=16\) reciprocal grid. The corresponding lists contain 8192 and 4096 full-zone points, respectively, and represent the same physical sampling after folding. The basis used \(R_{\rm MT}K_{\max}=7.0\) and \(G_{\max}=16~{\rm bohr}^{-1}\). Fixed muffin-tin radii were used within each elemental scan: 2.0 bohr for Li, 2.4 bohr for Na, and 2.5 bohr for K, Rb, and Cs. Total-energy and charge convergence thresholds were \(10^{-8}\) Ry and \(10^{-7}e\), respectively. Exact-folded \(N=24\) meshes independently verified representative onset points. The relevant energy gain is of order \(10^{-2}\)--\(10^{-1}\) meV/atom near onset.

The operational onset in Table~\ref{tab:si-bcc-onset} uses one common energy-resolution threshold,
\begin{equation}
E_{\rm sym}-E_{\rm free}>0.03~\mathrm{meV/atom},
\label{eq:si-operational-onset}
\end{equation}
which is applied uniformly across the five BCC alkali elements.

For the hP4 calculations, the low-pressure charge-equivalent branch and the high-pressure charge-transfer branch were analyzed separately. Self-consistent meshes were generated with a target of 2000 points in the full Brillouin zone and refined to 2500 points near the transition. We used \(R_{\rm MT}K_{\max}=7.0\), \(G_{\max}=16~{\rm bohr}^{-1}\), and energy and charge thresholds of \(10^{-8}\) Ry and \(10^{-7}e\), respectively. Pressure labels for the Na reference trajectory were obtained from an equation-of-state fit to the calculated \(E(V)\) data \cite{ref19,ref20}.

For the cross-element structural search, each element was sampled on a volume--\(c/a\) grid that was refined around the branch switch. At a specified external pressure, the structure was selected by minimizing \(H=E+PV\) over the converged grid. The Li, Na, and K results reported below follow these optimized hP4 structural paths.

Site charges \(Q_i\) were taken from the WIEN2k \texttt{:NTO} output, which reports the all-electron charge integrated inside the muffin-tin sphere of atom \(i\). The same muffin-tin radius and basis settings were used for the two sites and for the branches compared at each fixed geometry. For Type I, the diagnostic is the sublattice imbalance \(\Delta Q=Q_A-Q_B\). For Type II, it is \(\Delta Q=Q_{X_1}-Q_{X_2}\), where \(X_1=(0,0,0)\) and \(X_2=(1/3,2/3,1/4)\) in the ABAC setting. The charge-equivalent branch has \(\Delta Q=0\) to the converged precision of this partition, whereas the charge-transfer branch develops a reproducible finite \(\Delta Q\) together with the corresponding energy gain or band splitting.

\section{Landau Charge-Transfer Model}

Consider two sites \(A\) and \(B\) in a unit cell and define the charge-transfer order parameter \(\delta\) by
\begin{equation}
Q_A=Q_0+\delta e,\qquad Q_B=Q_0-\delta e .
\label{eq:si_delta}
\end{equation}
The charge-equivalent state corresponds to \(\delta=0\). If \(x\) exchanges \(A\) and \(B\), then
\begin{equation}
x:\delta\longmapsto-\delta .
\label{eq:si-delta-transformation}
\end{equation}
Whenever \(x\) belongs to the symmetry of the charge-equivalent problem, either as a crystallographic operation or as a hidden operation of the projected manifold, \(F(\delta)=F(-\delta)\). The free-energy change relative to this state is separated into an onsite charging cost and an intersite electrostatic contribution,
\begin{equation}
\Delta E(\delta,P)=\Delta E_{\rm intra}(\delta,P)+\Delta E_{\rm inter}(\delta,P).
\label{eq:si_energy_split}
\end{equation}
Near \(\delta=0\), Eq.~\eqref{eq:si-delta-transformation} requires an even expansion. Keeping the leading stabilizing term gives
\begin{align}
\Delta E_{\rm intra}(\delta,P)&=A_{\rm intra}(P)\delta^2+B_{\rm intra}(P)\delta^4,\\
\Delta E_{\rm inter}(\delta,P)&=-A_{\rm inter}(P)\delta^2 .
\end{align}
Thus
\begin{equation}
F_{\rm cell}(\delta,P)=A_2(P)\delta^2+A_4(P)\delta^4,
\label{eq:si_landau}
\end{equation}
where
\begin{align}
A_2(P)&=A_{\rm intra}(P)-A_{\rm inter}(P),\\
A_4(P)&=B_{\rm intra}(P)>0 .
\end{align}
The stationary points follow from
\begin{equation}
\frac{\partial F_{\rm cell}}{\partial\delta}
=2\delta\left[A_2(P)+2A_4(P)\delta^2\right]=0.
\label{eq:si-stationary}
\end{equation}
At \(\delta=0\), the curvature is \(\partial^2F_{\rm cell}/\partial\delta^2=2A_2\). The charge-equivalent state is stable when \(A_2(P)>0\). Its curvature vanishes at \(A_2(P)=0\), and the charge-transfer state is stable when
\begin{equation}
A_2(P)<0 .
\label{eq:si_instability}
\end{equation}
For \(A_2(P)<0\), the symmetry-broken state has
\begin{equation}
\delta=\pm\sqrt{\frac{-A_2(P)}{2A_4(P)}} .
\label{eq:si_delta_min}
\end{equation}
Substitution into Eq.~\eqref{eq:si_landau} gives
\begin{equation}
\Delta F_{\min}(P)=-\frac{A_2^2(P)}{4A_4(P)}.
\label{eq:si-energy-min}
\end{equation}
The order parameter grows as \((-A_2)^{1/2}\), while the energy gain grows as \(A_2^2\) below the instability.

\section{Electrostatic Estimate of the Instability}

The local charging cost can be estimated by treating the outer valence region as a spherical shell of effective radius \(R\), taken as slowly varying over the pressure interval relevant to the onset. The electrostatic self-energy of charging such a shell to \(Q\) is
\begin{equation}
E_{\rm self}=\frac{kQ^2}{2R},\qquad k=\frac{1}{4\pi\varepsilon_0}.
\end{equation}
For a two-site cell with charge transfer \(\delta\), the change in self-energy is
\begin{equation}
\Delta E_{\rm self}(\delta,P)=2\times\frac{k(\delta e)^2}{2R}
=\frac{ke^2}{R}\delta^2 .
\label{eq:si_self}
\end{equation}
This gives a classical estimate \(A_{\rm intra}\sim ke^2/R\).

Let the charge deviation on site \(i\) be \(q_i=s_i\delta e\), where \(s_i=\pm1\) labels the two charge sublattices. The electrostatic lattice energy per two-site cell is
\begin{equation}
E_{\rm lat}
=\frac{k(\delta e)^2}{2}
\sum_{\alpha\in\mathrm{cell}}
\sum_{j\ne\alpha}
\frac{s_\alpha s_j}{R_{\alpha j}}.
\label{eq:si-lattice-sum}
\end{equation}
The Madelung constant in this work is defined by
\begin{equation}
M=-\frac{r}{2}
\sum_{\alpha\in\mathrm{cell}}
\sum_{j\ne\alpha}
\frac{s_\alpha s_j}{R_{\alpha j}},
\label{eq:si-madelung-definition}
\end{equation}
where \(r\) is the nearest separation between opposite charge deviations. With this normalization, Eq.~\eqref{eq:si-lattice-sum} becomes
\begin{equation}
\Delta E_{\rm inter}(\delta,P)=-\frac{ke^2M(P)}{r(P)}\delta^2 .
\label{eq:si_lattice}
\end{equation}
The conditionally convergent sum in Eq.~\eqref{eq:si-madelung-definition} is evaluated by the Ewald procedure specified below for hP4. Combining Eqs.~\eqref{eq:si_self} and \eqref{eq:si_lattice} gives
\begin{equation}
A_2(P)\simeq ke^2\left[\frac{1}{R}-\frac{M(P)}{r(P)}\right].
\label{eq:si-a2-electrostatic}
\end{equation}
The charge-equivalent state therefore becomes unstable when
\begin{equation}
\frac{M(P)}{r(P)}>\frac{1}{R}
\quad \Longleftrightarrow \quad
r(P)<M(P)R .
\label{eq:si_geometric_criterion}
\end{equation}
This is the geometric criterion used in the main text. The derivation fixes the normalization of \(M\), \(r\), and \(R\) used in both structural classes.

\section{BCC Alkali-Metal Verification of the Geometric Criterion}

The Type-I group reduction can be stated exactly for BCC. The atomic structure has
\begin{equation}
G_{\rm lat}=Im\bar{3}m
\label{eq:si-bcc-glat}
\end{equation}
and contains the body-centering translation
\begin{equation}
t_c=\left\{E\,\middle|\,\frac12,\frac12,\frac12\right\}.
\label{eq:si-bcc-centering}
\end{equation}
This operation exchanges the two sublattices, so \(t_c:\Delta Q\mapsto-\Delta Q\). A selected state with \(\Delta Q\neq0\) therefore excludes \(t_c\). For the observed CsCl-type charge pattern, all operations that preserve each charge sublattice form
\begin{equation}
G=Pm\bar{3}m.
\label{eq:si-bcc-selected-group}
\end{equation}
The reduction is index two:
\begin{equation}
G_{\rm lat}=G\cup t_cG,\qquad
G=Pm\bar{3}m\subset G_{\rm lat}=Im\bar{3}m.
\label{eq:si-bcc-coset}
\end{equation}
The two degenerate domains \(\Delta Q=\pm\Delta Q_0\) are exchanged by \(t_c\). Thus the atomic BCC structure retains \(G_{\rm lat}\), while either spontaneously selected charge-transfer state realizes \(G\).

The relation between the translation and the site charges follows directly from the one-particle density matrix \(\rho\). Let \(P_A\) and \(P_B\) project onto the two site-centered sectors, with \(P_B=t_cP_At_c^{-1}\). If the electronic state retains \(t_c\), then \(t_c\rho t_c^{-1}=\rho\), and
\begin{equation}
Q_B=\operatorname{Tr}(P_B\rho)
=\operatorname{Tr}(P_A\rho)=Q_A.
\label{eq:si-bcc-charge-equality}
\end{equation}
A finite \(\Delta Q=Q_A-Q_B\) therefore measures the loss of \(t_c\) from the symmetry of the selected state.

For BCC, the nearest-neighbor separation and atomic volume are
\begin{equation}
r=\frac{\sqrt{3}}{2}a,\qquad V_{\rm atom}=\frac{a^3}{2}.
\end{equation}
For an alternating two-sublattice charge pattern, the Madelung constant is that of the CsCl structure, \(M_{\rm BCC}\simeq 1.762675\) \cite{ref16,ref26,ref27}. Taking \(R\) from empirical covalent-radius scales \cite{ref14,ref15}, the geometric criterion gives
\begin{equation}
r_c=M_{\rm BCC}R,\qquad
a_c=\frac{2M_{\rm BCC}R}{\sqrt{3}}.
\label{eq:si-bcc-critical-geometry}
\end{equation}
Using the common energy-resolution threshold and the folded-equivalent k meshes specified above, we extract the first-principles onset distances reported in Table~\ref{tab:si-bcc-onset}. For Li, Na, K, Rb, and Cs, the calculated \(r_{\rm DFT}\) values differ from the geometric predictions by no more than about \(6\%\).

This quantitative agreement across five alkali metals demonstrates that \(r_c=M_{\rm BCC}R\) captures the controlling structural scale of the Type-I instability without element-specific fitting.

\begin{table}[t]
\centering
\caption{Cross-element BCC onset distances for Type-I charge-symmetry breaking. Here \(r_c=M_{\rm BCC}R\) is the geometric prediction, while \(r_{\rm DFT}\) is the first-principles onset extracted with the common energy-resolution criterion in Eq.~\eqref{eq:si-operational-onset}.}
\label{tab:si-bcc-onset}
\footnotesize
\setlength{\tabcolsep}{3.2pt}
\begin{ruledtabular}
\begin{tabular}{lcccc}
Element & \(r_c\) & \(r_{\rm DFT}\) & \(\Delta r\) & \(\Delta r/r_c\)\\
 & (\(\text{\AA}\)) & (\(\text{\AA}\)) & (\(\text{\AA}\)) & (\%)\\
\hline
Li & 2.346 & 2.429 & \(+0.082\) & \(+3.5\)\\
Na & 2.731 & 2.841 & \(+0.110\) & \(+4.0\)\\
K  & 3.455 & 3.254 & \(-0.202\) & \(-5.8\)\\
Rb & 3.703 & 3.529 & \(-0.174\) & \(-4.7\)\\
Cs & 4.088 & 3.987 & \(-0.101\) & \(-2.5\)\\
\end{tabular}
\end{ruledtabular}
\end{table}

\section{Nonsymmorphic Degeneracy in the AB Parent Structure}

The Type-II construction begins with the AB (hcp) primitive cell of space group \(P6_3/mmc\) (No. 194). The boundary plane \(k_z=\pi/c\) contains the loop A--L--H--A, with
\begin{equation}
A=\left(0,0,\frac{1}{2}\right),\quad
L=\left(\frac{1}{2},0,\frac{1}{2}\right),\quad
H=\left(\frac{1}{3},\frac{1}{3},\frac{1}{2}\right),
\end{equation}
in fractional coordinates of the primitive reciprocal basis.

The relevant nonsymmorphic operation is the two-fold screw
\begin{equation}
S_{2z}\equiv \{C_{2z}|c/2\}.
\end{equation}
In the spinless case relevant for the present alkali-metal calculations,
\begin{equation}
S_{2z}^2=\{E|c\}.
\end{equation}
Acting on a Bloch state, \(\{E|c\}\) gives the phase \(e^{-ik_zc}\). Therefore on \(k_z=\pi/c\),
\begin{equation}
S_{2z}^2=-1.
\end{equation}
Combining the screw operation with time reversal \(\mathcal T\), define the antiunitary operation
\begin{equation}
\Theta=S_{2z}\mathcal T .
\end{equation}
For spinless electrons, \(\mathcal T^2=+1\) and \(\mathcal T\) commutes with spatial operations. On the \(k_z=\pi/c\) plane, \(\Theta\) leaves each momentum invariant modulo a reciprocal lattice vector. Therefore
\begin{equation}
\Theta^2=S_{2z}^2\mathcal T^2=\{E|c\},
\end{equation}
which acts on a Bloch state with the phase \(e^{-ik_zc}\). On the entire \(k_z=\pi/c\) plane,
\begin{equation}
\Theta^2=-1 .
\end{equation}
For an eigenstate \(|u_k\rangle\), antiunitarity gives
\begin{equation}
\langle\Theta u_k|\Theta^2u_k\rangle
=\langle\Theta u_k|u_k\rangle.
\label{eq:si-antiunitary-overlap}
\end{equation}
Using \(\Theta^2=-1\), the left-hand side equals \(-\langle\Theta u_k|u_k\rangle\). Hence \(\langle\Theta u_k|u_k\rangle=0\). The states \(|u_k\rangle\) and \(\Theta|u_k\rangle\) are orthogonal and degenerate at every momentum on the boundary plane. This gives the band sticking along A--L--H--A in the AB band structure.

\section{Folding from AB to ABAB}

Construct an ABAB supercell by doubling the primitive lattice along \(\hat z\), so that \(c'=2c\) and \(b_3'=b_3/2\). Momenta are identified modulo \(b_3'\). Expressed in the reciprocal basis of the doubled cell, the three vertices of the AB boundary loop fold as
\begin{equation}
\begin{aligned}
A=\left(0,0,\frac{1}{2}\right)
&\mapsto (0,0,1)\equiv \Gamma,\\
L=\left(\frac{1}{2},0,\frac{1}{2}\right)
&\mapsto \left(\frac{1}{2},0,1\right)\equiv M,\\
H=\left(\frac{1}{3},\frac{1}{3},\frac{1}{2}\right)
&\mapsto \left(\frac{1}{3},\frac{1}{3},1\right)\equiv K.
\end{aligned}
\end{equation}
Thus the full AB boundary loop A--L--H--A maps onto the basal loop \(\Gamma\)--M--K--\(\Gamma\) of the ABAB supercell. The nonsymmorphic degeneracy inherited from the AB parent structure is therefore transplanted onto the basal path of the ABAB cell.

Equivalently, introduce the folding vector
\begin{equation}
Q=b_3'=\frac{1}{2}b_3 .
\end{equation}
A supercell Bloch state at momentum \(k'\) contains components inherited from the two primitive-cell sectors \(k'\) and \(k'+Q\). The primitive translation \(T_c=\{E|c\}\) remains a symmetry of the ABAB reference structure. Its eigenvalues distinguish the two folded sectors,
\begin{align}
T_c|u_{n,k'}\rangle
&=e^{-ik_z'c}|u_{n,k'}\rangle,\nonumber\\
T_c|u_{m,k'+Q}\rangle
&=-e^{-ik_z'c}|u_{m,k'+Q}\rangle.
\label{eq:si-folded-sector-labels}
\end{align}
The two sectors cannot hybridize while \(T_c\) is retained. The boundary doublets of the primitive AB cell therefore remain degenerate after folding. The ABAB doublets along \(\Gamma\)--M--K--\(\Gamma\) are folded images of the nonsymmorphic-enforced boundary degeneracy of the AB parent structure.

\section{Hidden Gauge Equivalence between ABAB and ABAC}

The hP4 (ABAC) structure is not a simple \(2c\) supercell of AB. Nevertheless, at low pressure its basal-plane dispersion can closely mimic that of ABAB. The starting point is the hidden gauge equivalence established for an \(s\)-orbital-only close-packed model \cite{ref10}. We show below that the same construction extends to the short-range \((s,p_z)\) manifold relevant here. The group extension can be formulated by first identifying an ordinary ABAB space-group operation and then transporting it to the ABAC low-energy problem by the stacking gauge transformation.

In the four-layer ABAB cell, define
\begin{equation}
X_1=(0,0,0),\qquad
X_2=\left(\frac13,\frac23,\frac14\right).
\label{eq:si-abab-sites}
\end{equation}
Their translated partners are
\begin{equation}
X_1'=\left(0,0,\frac12\right),\qquad
X_2'=\left(\frac13,\frac23,\frac34\right).
\label{eq:si-abab-translated-sites}
\end{equation}
The operation
\begin{equation}
g=\left\{-I\,\middle|\,\frac13,\frac23,\frac14\right\}
\in G_{\rm lat}^{\rm ABAB}
\label{eq:si-abab-exchange}
\end{equation}
uses Seitz notation, where \(I\) is the three-dimensional identity matrix and the translation is expressed in fractional coordinates of the four-layer cell. It satisfies
\begin{equation}
\begin{aligned}
gX_1&=X_2, &\qquad gX_2&=X_1,\\
gX_1'&=X_2', & gX_2'&=X_1'
\end{aligned}
\qquad (\mathrm{mod}\,\mathbf R).
\label{eq:si-abab-four-site-map}
\end{equation}
It therefore places the two sublattices in one symmetry orbit and enforces their charge equivalence in ABAB. In ABAC, the fourth-layer site instead lies at
\begin{equation}
X_{2,\rm ABAC}'=\left(\frac23,\frac13,\frac34\right).
\label{eq:si-abac-fourth-site}
\end{equation}
Because
\begin{equation}
gX_1'=\left(\frac13,\frac23,\frac34\right)
\neq X_{2,\rm ABAC}'\quad(\mathrm{mod}\,\mathbf R),
\label{eq:si-g-fails-abac}
\end{equation}
the same real-space operation is absent from the ABAC crystallographic group:
\begin{equation}
g\notin G_{\rm lat}^{\rm ABAC}.
\label{eq:si-g-not-abac}
\end{equation}

Let a close-packed stacking be encoded by a chirality sequence
\begin{equation}
\sigma=\{\sigma_j=\pm1\},
\end{equation}
where \(\sigma_j\) specifies whether the stacking step from layer \(j\) to \(j+1\) turns right or left. In a layer-resolved basis, the Bloch Hamiltonian has the block form
\begin{equation}
H_{\sigma}(k)=
\begin{bmatrix}
D(k) & T_{\sigma_1}(k) & 0 & \cdots & T_{\sigma_M}^{\dagger}(k)\\
T_{\sigma_1}^{\dagger}(k) & D(k) & T_{\sigma_2}(k) & \cdots & 0\\
0 & T_{\sigma_2}^{\dagger}(k) & D(k) & \cdots & 0\\
\vdots & \vdots & \vdots & \ddots & \vdots\\
T_{\sigma_M}(k) & 0 & 0 & \cdots & D(k)
\end{bmatrix}.
\label{eq:si_block_hamiltonian}
\end{equation}
Here \(D(k)\) is the intralayer block, and \(T_{\pm}(k)\) are the interlayer hopping blocks for the two chiralities.

For a bond vector \(\eta\) with direction cosines \((l,m,n)\), the Slater--Koster hopping matrix in the \((s,p_z)\) subspace is
\begin{equation}
M^{(r)}(n)=
\begin{bmatrix}
V_{ss\sigma}^{(r)} & nV_{sp\sigma}^{(r)}\\
-nV_{sp\sigma}^{(r)} & n^2V_{pp\sigma}^{(r)}+(1-n^2)V_{pp\pi}^{(r)}
\end{bmatrix},
\label{eq:si_sk}
\end{equation}
where the superscript \(r\) labels the neighbor shell. This matrix depends on the out-of-plane direction cosine \(n\), but not on the in-plane direction of the bond. Let \(\{\eta_{-,\alpha}\}\) and \(\{\eta_{+,\alpha}\}\) denote the interlayer-neighbor bond sets for the two stacking chiralities. Close-packed geometry provides a one-to-one correspondence
\begin{equation}
\eta_{+,\alpha}=\eta_{-,\alpha}+s,
\qquad s_z=0,
\label{eq:si-bond-set-map}
\end{equation}
with the same out-of-plane direction cosine \(n_\alpha\) for each paired bond. The corresponding Bloch sums are therefore
\begin{align}
T_+(k)
&=\sum_\alpha e^{ik\cdot\eta_{+,\alpha}}M^{(r_\alpha)}(n_\alpha)\\
&=e^{ik\cdot s}\sum_\alpha e^{ik\cdot\eta_{-,\alpha}}M^{(r_\alpha)}(n_\alpha)
=e^{ik\cdot s}T_-(k),
\label{eq:si_tpm_phase}
\end{align}
Opposite stacking chiralities differ by a momentum-dependent phase within the short-range \((s,p_z)\) truncation. A chirality-independent hopping block is defined by
\begin{equation}
2\phi(k)=k\cdot s,
\qquad
\overline T(k)=e^{-i\phi(k)}T_+(k)=e^{i\phi(k)}T_-(k).
\label{eq:si-centered-hopping}
\end{equation}
The two blocks can then be written as
\begin{equation}
T_{\sigma_j}(k)=e^{i\sigma_j\phi(k)}\overline T(k).
\label{eq:si-chiral-block}
\end{equation}

For two stacking sequences \(\sigma\) and \(\chi\), choose a diagonal layer gauge
\begin{equation}
W_{\sigma,\chi}(k)=\mathrm{diag}
\left(e^{i\theta_1(k)},e^{i\theta_2(k)},\ldots,e^{i\theta_M(k)}\right).
\end{equation}
The off-diagonal blocks transform as
\begin{equation}
T_{\sigma_j}(k)\mapsto
e^{i(\theta_{j+1}-\theta_j)}T_{\sigma_j}(k).
\end{equation}
Using Eq.~\eqref{eq:si-chiral-block}, equality with \(T_{\chi_j}\) requires
\begin{equation}
\theta_{j+1}(k)-\theta_j(k)=(\chi_j-\sigma_j)\phi(k).
\label{eq:si-layer-phase-condition}
\end{equation}
With periodic boundary conditions, a solution exists when
\begin{equation}
\sum_{j=1}^{M}\sigma_j=\sum_{j=1}^{M}\chi_j .
\label{eq:si_equal_chirality}
\end{equation}
Indeed, summing Eq.~\eqref{eq:si-layer-phase-condition} over the layer index gives \(\theta_{M+1}-\theta_1=0\) when the two net chiralities are equal. The periodic identification \(\theta_{M+1}=\theta_1\) is then satisfied.
Under this equal-net-chirality condition,
\begin{equation}
W_{\sigma,\chi}^{\dagger}(k)H_{\sigma}(k)W_{\sigma,\chi}(k)=H_{\chi}(k).
\label{eq:si_unitary_equivalence}
\end{equation}
Thus the two stackings are isospectral within the truncated low-energy manifold.

For ABAB and ABAC, the relevant chirality sequences can be written as \(\sigma=(1,-1,1,-1)\) and \(\chi=(1,-1,-1,1)\). Both have zero net chirality. Choosing \(\theta_1=\theta_2=\theta_3=0\) and \(\theta_4=-k\cdot s\) gives the explicit ABAB-to-ABAC gauge map
\begin{equation}
U_g(k)=\operatorname{diag}\left(1,1,1,e^{-ik\cdot s}\right)\otimes I_2,
\label{eq:si-explicit-gauge-map}
\end{equation}
where \(I_2\) acts in the \((s,p_z)\) orbital subspace of each layer. It satisfies
\begin{equation}
U_g^\dagger(k)H_{\rm ABAB}(k)U_g(k)=H_{\rm ABAC}(k)
\label{eq:si-explicit-abab-abac-map}
\end{equation}
within the short-range, \((s,p_z)\)-dominated limit. The ABAB site exchange then induces
\begin{equation}
\widetilde g(k)=U_g^\dagger(gk)\,D_g(k)\,U_g(k),
\label{eq:si-gauge-conjugated-operation}
\end{equation}
where \(D_g(k)\) represents \(g\) in the folded ABAB Bloch basis. Using Eq.~\eqref{eq:si_unitary_equivalence} and the ABAB symmetry relation gives
\begin{equation}
\widetilde g(k)H_{\rm ABAC}(k)\widetilde g^\dagger(k)
=H_{\rm ABAC}(gk).
\label{eq:si-hidden-symmetry-relation}
\end{equation}
The relation follows explicitly from
\begin{align}
&\widetilde g(k)H_{\rm ABAC}(k)\widetilde g^\dagger(k)\nonumber\\
&=U_g^\dagger(gk)D_g(k)H_{\rm ABAB}(k)D_g^\dagger(k)U_g(gk)\nonumber\\
&=U_g^\dagger(gk)H_{\rm ABAB}(gk)U_g(gk)
=H_{\rm ABAC}(gk).
\label{eq:si-hidden-symmetry-derivation}
\end{align}
Thus \(\widetilde g\) is a symmetry of the projected ABAC Hamiltonian even though \(g\notin G_{\rm lat}^{\rm ABAC}\). More generally, applying the same gauge conjugation to every symmetry of the folded four-layer ABAB problem gives
\begin{equation}
G=U_g^\dagger G_{\rm fold}^{\rm ABAB}U_g
\supset G_{\rm lat}^{\rm ABAC},
\qquad
\widetilde g\in G\setminus G_{\rm lat}^{\rm ABAC}.
\label{eq:si-class-II-extension}
\end{equation}
Here \(G_{\rm fold}^{\rm ABAB}\) denotes the symmetry representation of the folded four-layer ABAB problem, and all groups in Eq.~\eqref{eq:si-class-II-extension} are understood through their representations projected onto the truncated \((s,p_z)\) manifold. The compact group conjugation is understood in the momentum-dependent sense made explicit in Eq.~\eqref{eq:si-gauge-conjugated-operation}. The full group relation explains why ABAC inherits the ABAB-like basal doublets, while the specific added operation \(\widetilde g\) explains why its two crystallographically inequivalent sites can remain electronically equivalent in that manifold.

The charge constraint can be stated with the site projectors. Let \(P_1(k)\) and \(P_2(k)\) project onto the \(X_1\) and \(X_2\) sectors of the ABAC low-energy basis. Since \(D_g\) exchanges the corresponding ABAB sectors, gauge conjugation gives
\begin{equation}
\widetilde g(k)P_1(k)\widetilde g^\dagger(k)=P_2(gk).
\label{eq:si-hidden-projector-map}
\end{equation}
For a density matrix that retains \(\widetilde g\), the projected site charges are
\begin{equation}
Q_i=\sum_k\operatorname{Tr}\left[P_i(k)\rho(k)\right].
\label{eq:si-projected-charge}
\end{equation}
Using Eq.~\eqref{eq:si-hidden-projector-map}, the symmetry of \(\rho\), and a change of summation variable from \(k\) to \(gk\) gives
\begin{equation}
Q_2=\sum_k\operatorname{Tr}\left[P_2(k)\rho(k)\right]
=\sum_k\operatorname{Tr}\left[P_1(k)\rho(k)\right]=Q_1.
\label{eq:si-hidden-charge-equality}
\end{equation}
The hidden operation therefore enforces charge equality inside the projected manifold. A finite \(\Delta Q=Q_1-Q_2\) is odd under \(\widetilde g\) and removes this operation from the symmetry of the selected state.

\begin{figure}[t]
\centering
\includegraphics[width=0.92\textwidth]{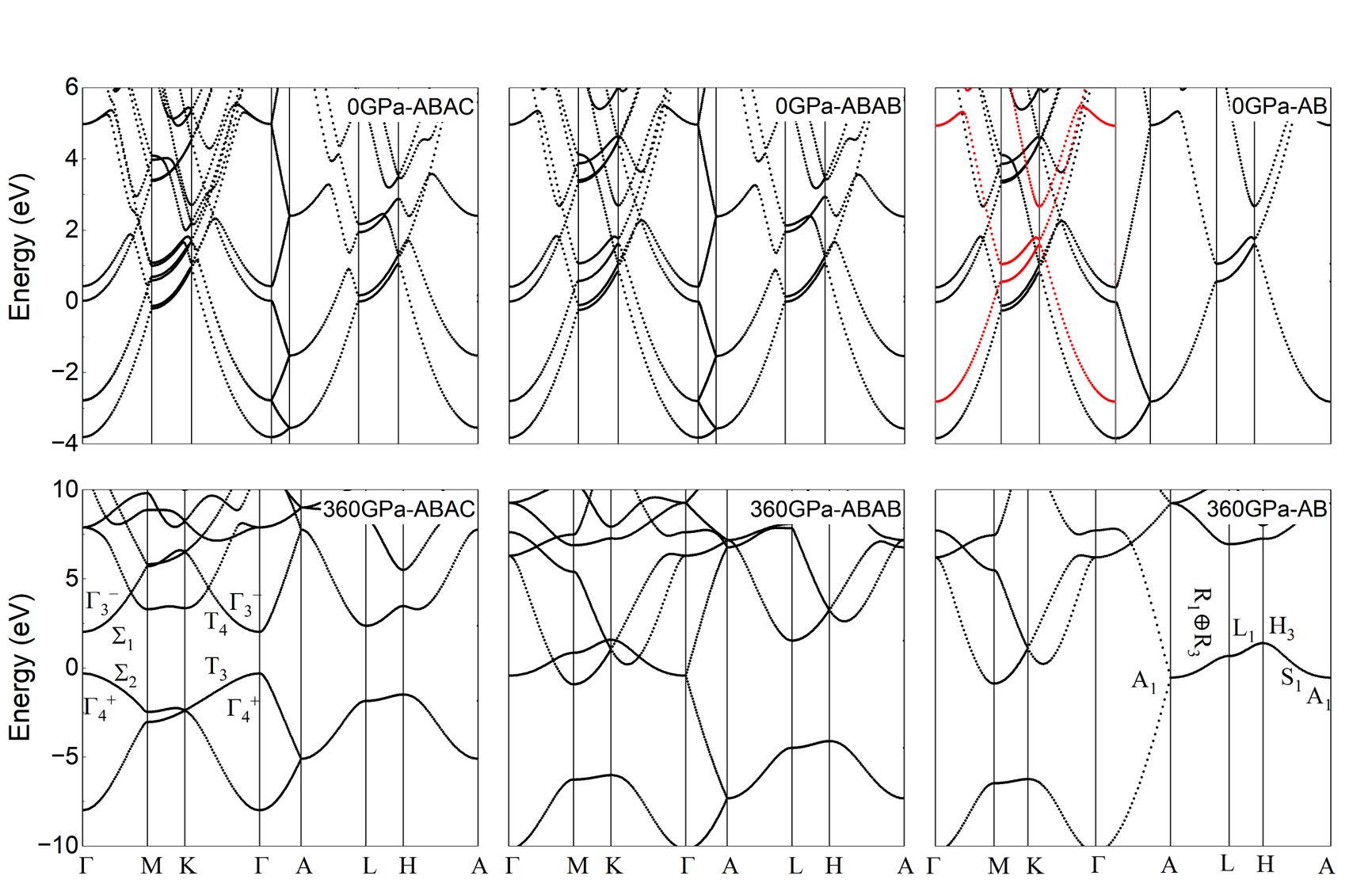}
\caption{Band-level origin of the hidden charge symmetry. At low pressure, ABAC and the folded ABAB reference show nearly identical basal-plane dispersions. The ABAB doublets are folded images of the nonsymmorphic band sticking on A--L--H--A in the AB parent. Under compression, the ABAC and ABAB spectra separate as the low-energy gauge mapping fails and the charge-transfer imbalance removes the hidden site exchange \(\widetilde g\).}
\label{fig:si-stacking-comparison}
\end{figure}

\section{Pressure-Induced Breakdown of Hidden Equivalence}

The hidden equivalence is exact in the truncated short-range \((s,p_z)\) model. Write the compressed low-energy Hamiltonian as
\begin{align}
H(k,P)={}&H_0(k,P)+\Delta H_{\rm orb}(k,P)\nonumber\\
&+\Delta H_{\rm lr}(k,P)+\Delta H_Q(P).
\label{eq:si-compressed-hamiltonian}
\end{align}
The reference term \(H_0\) satisfies Eq.~\eqref{eq:si-hidden-symmetry-relation}. The term \(\Delta H_{\rm orb}\) contains the increasing \(p_{x,y}\) and \(d\) components. Their Slater--Koster matrix elements depend on the in-plane direction cosines \(l\) and \(m\), so opposite stacking chiralities are not related by a single phase. The term \(\Delta H_{\rm lr}\) contains longer-range interlayer processes whose bond sets are not related by the uniform shift in Eq.~\eqref{eq:si-bond-set-map}. These terms invalidate Eq.~\eqref{eq:si_tpm_phase} and progressively separate the ABAB and ABAC spectra.

The charge-transfer contribution is diagonal in the projected site sectors and can be written as
\begin{equation}
\Delta H_Q(P)=\frac{\Delta\varepsilon(P)}{2}
\left[P_1-P_2\right].
\label{eq:si-charge-transfer-perturbation}
\end{equation}
Equation~\eqref{eq:si-hidden-projector-map} gives
\begin{equation}
\widetilde g\,\Delta H_Q\,\widetilde g^\dagger=-\Delta H_Q.
\label{eq:si-charge-transfer-odd}
\end{equation}
A finite \(\Delta\varepsilon\), accompanied by \(\Delta Q=Q_{X_1}-Q_{X_2}\ne0\), therefore removes \(\widetilde g\) from the symmetry of the selected state. The near-Fermi partners along K--\(\Gamma\)--M belong to different one-dimensional irreducible representations of the corresponding little groups. A perturbation that preserves those little groups has no off-diagonal matrix element between the partners. The loss of hidden equivalence therefore produces a direct energy separation rather than an avoided crossing.

The geometric comparison in Table~\ref{tab:si-hp4-onset} uses the same criterion as the BCC analysis, but with the hP4 Madelung constant for the alternating \(X_1/X_2\) charge-transfer pattern. For the ABAC structure with \(X_1=(0,0,0)\) and \(X_2=(1/3,2/3,1/4)\), the nearest inter-sublattice distance is
\begin{equation}
r_{12}=\left(\frac{a^2}{3}+\frac{c^2}{16}\right)^{1/2}.
\label{eq:si-hp4-distance}
\end{equation}
The in-plane term follows from the hexagonal metric, \(|x\mathbf a_1+y\mathbf a_2|^2=a^2(x^2+y^2-xy)\), evaluated at \((x,y)=(1/3,2/3)\). The out-of-plane separation is \(c/4\).

The Madelung constants are evaluated by an Ewald sum for charges \(q_i=s_i\delta e\) in a neutral cell. With Ewald parameter \(\eta\), cell volume \(\Omega\), direct lattice vectors \(\mathbf R\), and reciprocal vectors \(\mathbf G\), the electrostatic energy per cell is
\begin{align}
E_{\rm Ewald}={}&\frac{k}{2}
\sum_{ij,\mathbf R}'q_iq_j
\frac{\operatorname{erfc}\!\left(\eta|\mathbf r_{ij}+\mathbf R|\right)}
{|\mathbf r_{ij}+\mathbf R|}\nonumber\\
&+\frac{2\pi k}{\Omega}
\sum_{\mathbf G\ne0}
\frac{e^{-G^2/(4\eta^2)}}{G^2}
\left|\sum_jq_je^{i\mathbf G\cdot\mathbf r_j}\right|^2
\nonumber\\
&-\frac{k\eta}{\sqrt\pi}\sum_jq_j^2.
\label{eq:si-ewald}
\end{align}
The prime omits the \(i=j\), \(\mathbf R=0\) self term. Charge neutrality removes the \(\mathbf G=0\) contribution. The Madelung constant is extracted from
\begin{equation}
M_{\rm hP4}=-\frac{r_{12}E_{\rm Ewald}}
{k(\delta e)^2}.
\label{eq:si-ewald-madelung}
\end{equation}
The converged value is independent of \(\eta\). The values used in the main-text comparison are listed in Table~\ref{tab:si_hp4_lattice_factors} and use the same normalization as \(M_{\rm hP4}(c/a=1.85288)=1.4298\).

\FloatBarrier

\begin{table}[t]
\centering
\caption{hP4 onset distances for the Type-II charge-transfer instability. Here \(r_c=M_{\rm hP4}R\), while \(r_{\rm DFT}\) is the nearest \(X_1\)--\(X_2\) distance at the first-principles transition where hidden charge equivalence breaks down.}
\label{tab:si-hp4-onset}
\footnotesize
\setlength{\tabcolsep}{3.2pt}
\begin{ruledtabular}
\begin{tabular}{lcccc}
Element & \(r_c\) & \(r_{\rm DFT}\) & \(\Delta r\) & \(\Delta r/r_c\)\\
 & (\(\text{\AA}\)) & (\(\text{\AA}\)) & (\(\text{\AA}\)) & (\%)\\
\hline
Li & 1.852 & 1.953 & \(+0.101\) & \(+5.4\)\\
Na & 2.216 & 2.154 & \(-0.062\) & \(-2.8\)\\
K  & 2.745 & 2.939 & \(+0.194\) & \(+7.1\)\\
\end{tabular}
\end{ruledtabular}
\end{table}

\begin{table}[t]
\centering
\caption{hP4 structural and electrostatic parameters used for the Type-II onset comparison in the main text. The listed \(c/a\) values specify the first-principles transition geometries, \(M_{\rm hP4}\) is the Madelung constant for the alternating \(X_1/X_2\) charge-transfer pattern, and \(R\) is the empirical charging radius used in \(r_c=M_{\rm hP4}R\).}
\label{tab:si_hp4_lattice_factors}
\footnotesize
\setlength{\tabcolsep}{4pt}
\begin{ruledtabular}
\begin{tabular}{lccc}
Element & \(c/a\) & \(M_{\rm hP4}\) & \(R\)\\
 & & & (\(\text{\AA}\))\\
\hline
Li & 2.20 & 1.393 & 1.33\\
Na & 1.85288 & 1.4298 & 1.55\\
K  & 1.40 & 1.400 & 1.96\\
\end{tabular}
\end{ruledtabular}
\end{table}

\FloatBarrier

Combining these lattice factors with the empirical radii gives \(r_c=M_{\rm hP4}R\). The resulting critical distances differ from the first-principles transition distances by \(5.4\%\), \(2.8\%\), and \(7.1\%\) for Li, Na, and K, respectively. The same geometric criterion therefore describes both the Type-I instability of space-group-equivalent sites and the Type-II breakdown of hidden charge equivalence.

\section{Irrep-Resolved Alkali-Metal Results}

Table~\ref{tab:class2_linak_irrep_splitting_si} summarizes the irrep labels and splittings of the two near-Fermi partner bands for Li, Na, and K. In all three elements, the low-pressure charge-equivalent branch contains the near-degenerate partner manifold, whereas charge transfer lifts the hidden equivalence and separates the partner bands. For K, the charge-transfer structure is chosen immediately beyond the transition and the partners are followed by their compatible irreducible representations through intervening crossings with other bands.

\begin{table}[t]
\centering
\scriptsize
\setlength{\tabcolsep}{3pt}
\caption{Irreducible representations and splittings of the two near-Fermi partner bands in hP4 (ABAC) Li, Na, and K. EQ denotes the low-pressure charge-equivalent branch, where the partner irreps are listed as a near-degenerate pair. CT denotes the charge-transfer branch, where irreps are listed as lower/upper branches in energy.}
\label{tab:class2_linak_irrep_splitting_si}
\begin{tabular}{ccccccc}
\hline
Element & Branch & \(V_{\rm atom}\) (\(\text{\AA}^3\)) & \(c/a\) & K--\(\Gamma\) & \(\Gamma\) & \(\Gamma\)--M \\
\hline
Li & EQ & 12.00 & 3.20 & \(T_3\oplus T_4\) (5.5 meV) & \(\Gamma_{4}^{+}\oplus \Gamma_{3}^{-}\) (0.1 meV) & \(\Sigma_2\oplus \Sigma_1\) (2.5 meV) \\
Li & CT & 7.50 & 2.50 & \(T_3/T_4\) (93.2 meV) & \(\Gamma_{4}^{+}/\Gamma_{3}^{-}\) (23.1 meV) & \(\Sigma_2/\Sigma_1\) (57.9 meV) \\
Na & EQ & 9.60 & 3.20 & \(T_3\oplus T_4\) (6.6 meV) & \(\Gamma_{3}^{-}\oplus \Gamma_{4}^{+}\) (0.0 meV) & \(\Sigma_2\oplus \Sigma_1\) (3.8 meV) \\
Na & CT & 8.80 & 1.70 & \(T_3/T_4\) (757.5 meV) & \(\Gamma_{4}^{+}/\Gamma_{3}^{-}\) (258.5 meV) & \(\Sigma_2/\Sigma_1\) (590.2 meV) \\
K & EQ & 27.00 & 3.20 & \(T_3\oplus T_4\) (18.7 meV) & \(\Gamma_{4}^{+}\oplus \Gamma_{3}^{-}\) (4.1 meV) & \(\Sigma_2\oplus \Sigma_1\) (12.5 meV) \\
K & CT & 25.00 & 1.40 & \(T_3/T_4\) (2016.9 meV) & \(\Gamma_{4}^{+}/\Gamma_{3}^{-}\) (950.2 meV) & \(\Sigma_2/\Sigma_1\) (1895.9 meV) \\
\hline
\end{tabular}
\end{table}

The irrep table verifies the same hidden-equivalence breakdown across Li, Na, and K. The irrep-compatible partners remain identifiable even when their energy ordering changes through crossings with other bands.

\begin{figure}[t]
\centering
\includegraphics[width=0.96\textwidth]{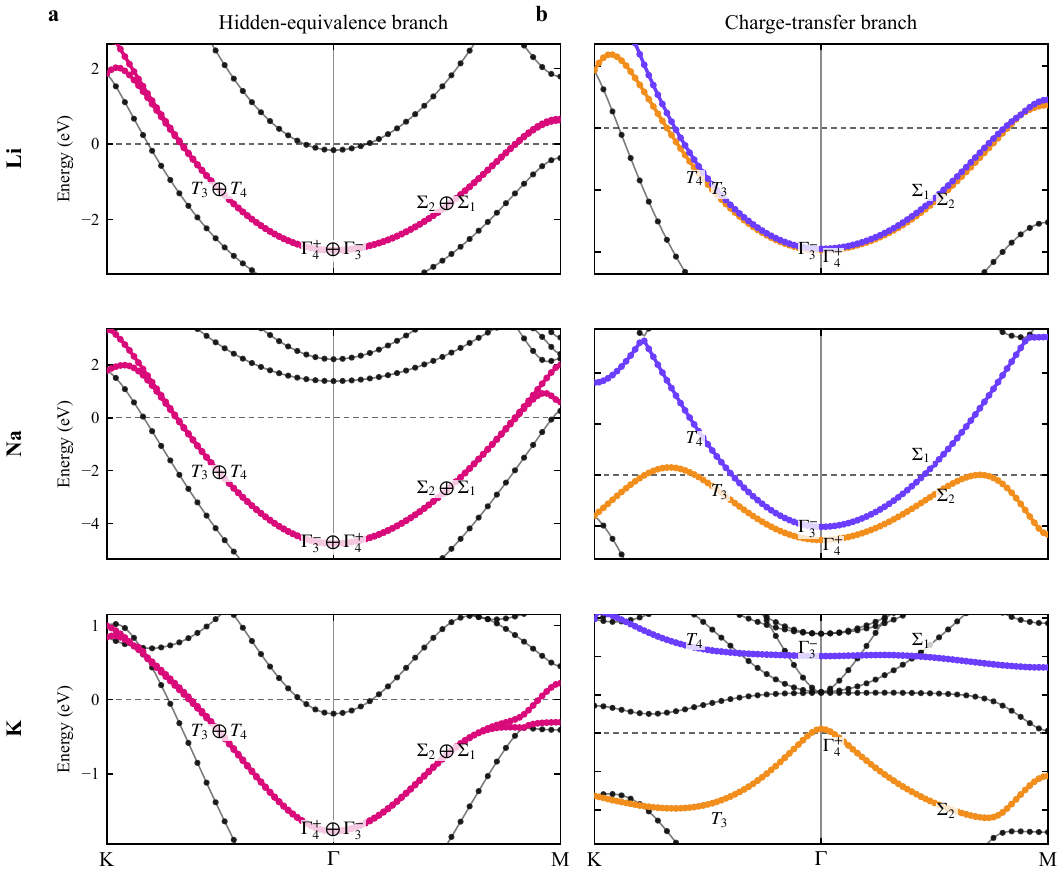}
\caption{Irrep-resolved Type-II results for hP4 alkali structures. Low-pressure hidden-equivalence (EQ) branches are shown on the left and charge-transfer (CT) branches on the right for Li, Na, and K. The magenta bands highlight the near-degenerate partner manifold in the EQ branch. Purple and orange mark the corresponding split branches after charge transfer. In the K panel, the CT structure is the first converged point beyond the transition, and the colored branches are tracked by the compatibility sequence \(T_3/T_4\rightarrow\Gamma_4^+/\Gamma_3^-\rightarrow\Sigma_2/\Sigma_1\) through allowed crossings with other bands.}
\label{fig:si-linak-bands}
\end{figure}

\FloatBarrier

\end{document}